\documentclass[aps,prl,twocolumn,showpacs]{revtex4}
\usepackage{epsfig}
\usepackage{graphicx}  
\usepackage{dcolumn}   
\usepackage{amssymb,amsmath}

\newcommand{\be}{\begin{equation}} \newcommand{\ee}{\end{equation}}
\newcommand{\bea}{\begin{eqnarray}} \newcommand{\eea}{\end{eqnarray}}

\begin{document}

\title{Coherent electron-phonon states in suspended quantum dots: decoherence and dissipation effects}
\author{Luis G. C. Rego}
\affiliation{Departamento de F\'{\i}sica,
Universidade Federal de Santa Catarina, Florian\'opolis, SC,
88040-900, Brazil}

\date{\today}

\begin{abstract}
The dynamics of coherent electron-phonon (el-ph) states is investigated for a suspended nanostructure.
Exact quantum dynamics calculations reveal that electron and phonons (comprising a thermal bath) couple quantum mechanically to perform coherent oscillations with periods in the range of tens of nanoseconds, despite the finite temperature of the phonon bath. 
Mechanical energy dissipation due to clamping loss is taken into account in the calculations.
Although the lifetime of the coupled el-ph states decreases with the temperature, well defined Rabi oscillations are obtained for temperatures up to 100 mK. The dynamics of the coupled electron-phonon state is susceptible to various forms of external control. For instance, a weak external magnetic field can be used to control the dynamics of the system, by decoupling the electron from the phonon bath. The results cast light upon the underlying physics of a yet unexplored system that could be suitable for novel quantum device applications.
\end{abstract}


\maketitle

\section{Introduction}

The possibility of engineering semiconductor devices at the nano and micro scales has created the conditions for testing fundamental aspects of quantum theory otherwise difficult to probe in natural atomic size systems. Particularly, quantum dot (QD) systems have been recognized as a physical realization of artificial atoms and molecules, whose properties (e.g., structural and transport) have been intensively investigated, notably in the presence of magnetic fields \cite{Pawel,Kouwenhoven,Beenakker,Kotthaus, Elzerman}. 
QDs have become the building blocks of various quantum devices and, nowadays, they can be coupled in arrays to create charge \cite{Gorman,Hayashi} or spin qubit \cite{Loss,Petta,Wiel} gates. 

A new promising possibility for both implementing and investigating
coherent phenomena in semiconductor devices is the combination of quantum dot systems with suspended nanostructures \cite{Brandes,Blick02,nano,Blencowe,Cleland,Gusso}. As an immediate consequence, it is possible to improve the isolation of the electronic quantum system from the bulk of the sample. But it is also expected that nanoelectromechanical systems (NEMS) will lead to the investigation of new regimes of phonon-mediated processes \cite{Brandes,Gusso} and the observation of quantum behavior in mesoscopic mechanical systems. In addition, phonon cavities can be envisaged as a solid state analog of quantum electrodynamic cavities.

For all these pursuits it is generally desirable to construct systems with very little loss of energy and very high quality factors Q. High frequency resonators have been fabricated from a variety of materials, most of them based on silicon.
Due to potential applications as on-chip high-Q filters and clocks, nanomechanical resonators are rapidly being pushed to smaller size scales and higher frequencies. Resonant frequencies higher than 1 GHz have consistently been achieved \cite{Huang,Gaidarzhy}. However, it has also been observed that the quality factor Q of micro and nano resonators decrease as the size of the system is reduced and the fundamental frequency increased \cite{Qfactor,Mohanty02,HuangNJP}. 
That effect imposes severe constraints on the design of NEMS oscillators, specially for those applications that rely on the response of a single resonant frequency.
Therefore, it is important to look for alternative systems that can preserve coherence even in the case of high mechanical dissipation.

\begin{figure}[h]
\includegraphics[width=6cm]{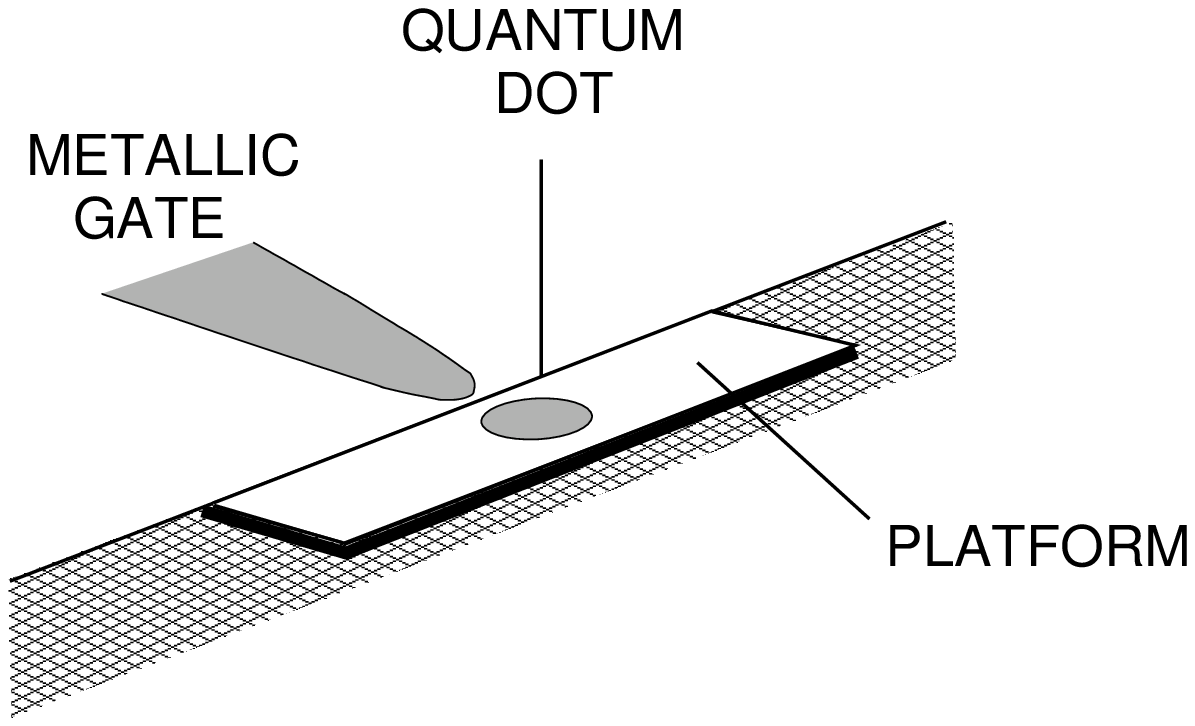}
\caption{Illustration of the NEMS structure comprised of a circular quantum dot of radius $R=75~nm$ embedded on a suspended platform with dimensions $w=1.2~\mu m$ (width), $l=200~nm$ (length) and $t=50~nm$ (thickness). The figure also depicts a metallic gate extending through the bulk of the sample, which affords external control of the state of the electron with minimal interference with the vibrational state of the platform.}
\label{structure}
\end{figure}

Here we investigate the underlying physics of a novel type of nanoelectromechanical oscillator that is based on the dynamics of coupled electron-phonon states. The work looks into the evolution of a QD state with well defined electronic angular momentum as it interacts with a thermal field comprised by the phonon states of the platform at finite temperature. As a consequence of the electron-phonon interaction, the electronic angular momentum oscillates between degenerate eigenstates of $\hat{L}_z$, like in a resonant Rabi system. A weak magnetic field can be used to decouple the electron from the thermal bath, by splitting the $\hat{L}_z$ eigenstates without disturbing the original symmetry of the electronic wavefunctions.

\begin{figure}
 \includegraphics[width=6cm]{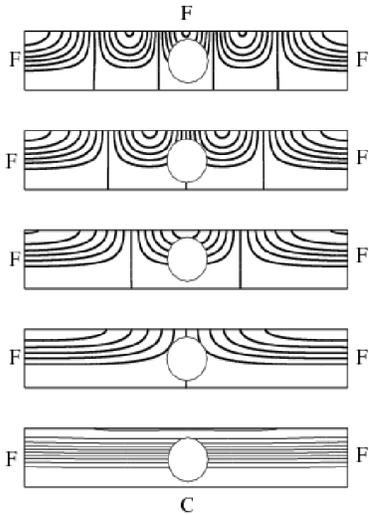}
\caption{Contour plots of the transverse vibrations of the platform. From the bottom to the top, the first 5 phonon modes are depicted. The phonon modes have even (+) or odd (-) parity with respect to the width dimension. The letters C and F denote the clamped and free edges. The circle represent the position of the QD.}
\label{mode-shapes}
\end{figure}

For that purpose,
we consider a yet unexplored design for the suspended NEMS, whose width ($w$) is larger than the length ($l$), as shown in Figure \ref{structure}. For the sake of convenience we name it platform, to make the distinction with the usual cantilever structures. The dimensions of the suspended platform used in the following calculations are $w=1.2~\mu m$, $l=200~nm$ and $t=50~nm$ (thickness).
Among the reasons that moved us to consider such NEMS, we point out that its {\it fundamental frequency} ($f_1 \approx 2~GHz$) is approximately 40 times higher than the fundamental frequency of a cantilever with equivalent dimensions ({\it i.e.}, $w=200~nm$, $l=1.2~\mu m$). The short spaced density of states of a normal cantilever (or bridge) favor dissipation and wash out the coherent electron-phonon dynamics.
In addition, the proposed configuration allows a direct external contact with the electron in the quantum dot, by means of an electrostatic gate extending 
through the bulk of the sample, as depicted in the Figure \ref{structure}. We notice that such a noninvasive external coupling with the QD is not possible in the regular cantilever or bridge structures. The lowest transverse vibrational modes of the platform are depicted in Figure \ref{mode-shapes} together with the circular QD.

Despite the advantages, the proposed NEMS poses significant caveats compared to the usual cantilever and bridge structures. The most serious difficulty is the augmented dissipation of vibrational energy through the wide clamping edge of the platform. However, we find that the deleterious effects caused by the low Q factor of the platform are compensated by the fact the dynamics of the QD is determined from outset by its interaction with a thermal field. The influence of external and internal mechanisms of energy dissipation will be discussed in the light of the simulation results.

Section \ref{Model} presents the theoretical methods developed to calculate the quantum dynamics of the proposed nanoelectromechanical structure at finite temperatures.
Section \ref{simulations} is dedicated to the analysis of the simulation results under different physical conditions. The conclusions and a final discussion are presented in section \ref{conclusions}.

\section{Theoretical Model}
\label{Model}

\subsection{Suspended electron-phonon system}
\label{model-A}

The hamiltonian of the system is written as
\bea
\hat{H} &=& \hat{H}_{el} + \hat{H}_{ph} + \hat{H}_{el-ph} + \hat{H}_{relax}
\nonumber \\
&=& \sum_\kappa E_\kappa b^\dagger_\kappa b_\kappa +
\sum_\alpha \left( \hat{n}_\alpha +\frac12 \right) \hbar \omega_\alpha
\nonumber \\ &+&
\sum_{\alpha \, \kappa \, \kappa'} 
g_{\kappa' \alpha \kappa}
\ b^\dag_{\kappa'}\left[a_{\alpha}^\dag + a_{\alpha}\right] b_{\kappa} \nonumber \\
&+& \sum_{\alpha}
i~\gamma_\alpha \left[\overline{n_\alpha}(T) - \langle \hat{n}_\alpha \rangle\right] \delta_{\kappa,\kappa'}\delta_{{\bf n},{\bf n'}} ,
\label{hamiltonian}
\eea
which includes the electron and phonon hamiltonians ($\hat{H}_0 = \hat{H}_{el} + \hat{H}_{ph}$), the electron-phonon interaction ($\hat{H}_{el-ph}$) and the phonon relaxation term ($\hat{H}_{relax}$). In brief,
$b^\dagger_\kappa$ and $b_\kappa$ are the electron creation and annihilation operators, while $a_{\alpha}^\dag$ and $a_{\alpha}$ are the corresponding phonon operators, with $\hat{n}_\alpha = a_{\alpha}^\dag a_{\alpha}$. The el-ph matrix elements are denoted by $g_{\kappa' \alpha \kappa}$. The energy relaxation rate of the phonon system with the external environment is $\gamma_\alpha = \hbar\omega_\alpha/Q$, where $Q$ is the quality factor of the platform, $\overline{n_\alpha}(T)$ is the thermal occupation of phonon mode $\alpha$ and $\langle \hat{n}_\alpha \rangle$ the corresponding quantum average for the coupled el-ph state.

Hamiltonian (\ref{hamiltonian}) is a generalization of the usual atom-field interaction \cite{Scully}, here describing a multi-level electronic system in a multi-mode bosonic cavity . In the following calculations we assume that the QD is occupied by a single electron -- such devices have been extensively characterized in the literature \cite{Kotthaus,JAP92,APL82,PRL94,NL5}. The model can also describe the excess electron of a QD.
The one electron hamiltonian for a quasi two-dimensional circular QD of radius $R$ has eigenstates
\begin{equation}
\varphi_{\kappa}(r,\theta) = \frac{{\mbox J}_{|l|}
\left(\alpha_{l\nu} \frac{r}{R}\right) \exp[i \, l \, \theta]}
{\sqrt{\pi}R|J_{|l|+1}(\alpha_{l\nu})|},
\label{pureelectron}
\end{equation}
with $\kappa \equiv (l, \nu)$, $l = 0, \pm 1, \pm 2,\ldots$ and
$\alpha_{l \nu}$ the $\nu$-th root of the Bessel function of order
$|l|$, $J_{|l|}(\alpha_{l\nu}x)$. The corresponding energies are  $E_{\kappa} = \frac{\hbar^2}{2 m_e} \frac{\alpha_{l \nu}^2}{R^2}$, where $m_e$ is the effective mass of the electron. 
An interesting condition arises if a weak magnetic field $\vec{B}$ is applied perpendicularly to the plane of the quantum dot \cite{Luz}. In this case the one electron hamiltonian becomes 
\begin{eqnarray}
\hat{H}_{el} &=& \frac{1}{2m_e}\left[ \mathbf{p} - \frac{e}{c}\vec{A} \right]^2 + V(r) 
\label{magnetic1} \\
&=&-\frac{\hbar^2}{2m_e}\nabla^2 + V(r) + \frac{eB}{2m_e c}~L_z + \frac{e^2B^2}{8m_e c^2}r^2
\label{magnetic2} 
\end{eqnarray}
where $V(r)$ is zero for $r<R$ and infinity outside the dot. The vector potential $\vec{A} = B/2~(-y,x,0)$ is written in the symmetric gauge. For weak magnetic fields, such that the magnetic length $l_B = \left(\hbar c/eB\right)^{1/2}$
is larger than the radius of the dot, $R<l_B$, the diamagnetic term ($E_{diam}=\frac{e^2 B^2}{8m_e c^2}r^2$) in Eq (\ref{magnetic2}) can be disregarded and the one electron energies can be written as 
\begin{eqnarray}
 E_{l\nu} = \frac{\hbar^2}{2 m_e} \frac{\alpha_{l \nu}^2}{R^2} + \mu_{_B}mB \ ,
\label{Zeeman}
\end{eqnarray}
where $\mu_{_B} = e\hbar/2m_e c$ is the Bohr magneton. Moreover, in the weak field approximation the wavefunctions (\ref{pureelectron}) are still eigenstates of $\hat{H}_{el}$, since $\hat{L}_z = -i\hbar\partial/\partial \theta$ commutes with the hamiltonian. For $R=75~nm$ and $B=500~G$, such that $l_B \simeq 1.5 R$, $E_{diam}/E_{Zeeman} = 0.1$ and the error committed by Eq. (\ref{Zeeman}) in the calculation of the energy is less than 1\% (more details in Section \ref{withB}).

The phonon eigenstates of the platform are obtained through the quantization of its mechanical modes of vibration, according to the formalism detailed in Ref [\onlinecite{Gusso}]. 
The operators $a^\dagger_\alpha$ and $a_\alpha$, associated to $\hat{n}_\alpha = a^\dagger_\alpha a_\alpha$ in Eq. (\ref{hamiltonian}), create and annihilate phonon modes in the cavity. At low temperatures only the long wavelength acoustic modes are relevant and the semiconductor is described by the elastic approximation, so that
the classical plate theory (CPT) \cite{Graff} can be used to calculate 
the vibrational modes of a thin platform. The {\it fundamental frequency} calculated for the suspended platform depicted in Fig. \ref{structure} is $f_1 \approx 1.88$ GHz. The dynamics of the coupled electron-phonon bath state depends on the spectrum of the phonon modes, therefore it is important to validate the calculated phonon energies. For that purpose we have compared the CPT modes with those obtained from the rigorous three-dimensional (3D) elastic equations \cite{Zhou,Gusso07}. It is found that the CPT method consistently yields the correct mode shapes and, moreover, the fundamental frequency $f_1$ evaluated by the CPT method is just 3\% higher than the value yielded by 3D method. It is also well known that the CPT method tends to overestimate the vibrational frequencies as the order of the modes increase, however, the energy discrepancy of the higher modes is compensated by their low thermal occupation for temperatures $T \lesssim 100$ mK.

The hamiltonian for the electron-phonon interaction reads
\begin{eqnarray} 
\hat{H}_{el-ph} = C_{DP} \sum_{\alpha \, \kappa' \kappa}
\frac{V_{\kappa' \, \alpha \, \kappa}^{DP}}{\sqrt{\omega_{\alpha}}}
\, b^\dag_{\kappa'} \left[a_{\alpha}^\dag + a_{\alpha}\right]
b_{\kappa} \ ,
\label{DP} 
\end{eqnarray} 
where $g_{\kappa' \, \alpha \, \kappa} =  C_{DP} V_{\kappa' \, \alpha \, \kappa}^{DP} / \sqrt{\omega_{\alpha}}$, in connection to Eq. (\ref{hamiltonian}), $C_{PD}$ denotes the deformation potential (DP) constant of the material and $b^\dagger_\kappa$ ($b_\kappa$) is the fermionic creation (annihilation) operator satisfying the usual anti-commutation relations.
The electron-phonon interaction is written in terms of the matrix elements $V_{\kappa'\alpha\kappa}$ that describe the coupling between electrons and phonons in the NEMS. The properties of $V_{\kappa'\alpha\kappa}$ depend on both the material attributes and the geometrical symmetries of the device. At low temperatures the more important electron-phonon coupling mechanisms are described by the deformation (DP) and the piezoelectric (PZ) potentials, however, only the former is included in (\ref{hamiltonian}), because silicon is not a piezoelectric material.
 A detailed derivation of Eq. (\ref{DP}), including the piezoelectric interaction, can be found in Ref [\onlinecite{Gusso}]. Furthermore, it has also been demonstrated \cite{Gusso} that, for a circular QD located at the center of the platform, the even parity phonon modes produce real valued matrix elements $g_{\kappa'\alpha \kappa}$ whereas the odd parity phonon modes yield imaginary valued elements. If the QD is displaced from the center of the platform  $g_{\kappa' \alpha \kappa}$ is complex. Moreover, the matrix representation of the total hamiltonian $\hat{H}$ can be written in blocks of fixed electronic angular momenta $l$'s as $\mathbb{H}_{l'l} = \mathbb{A}_{l'l} +
(i)^{\mbox{\scriptsize{sign}}(l' - l)} \, \mathbb{B}_{l'l}$, where $\mathbb{A}$ and $\mathbb{B}$ are real, symmetric and mutually disjoint matrices.
Taken individually, the matrix $\mathbb{A}$  ($\mathbb{B}$) is responsible for the coupling between the electron and the even (odd) parity phonon modes.

Finally, the dominant source of mechanical energy dissipation in the NEMS is due to attachment loss via the platform clamping edge. The mechanism is described within the relaxation time approximation in the hamiltonian (\ref{hamiltonian}), with $\gamma_\alpha = \hbar\omega_\alpha/Q$, where $Q$ is the quality factor of the NEMS, assumed to be the same for all the modes. In Eq. \ref{hamiltonian}, $\overline{n_\alpha}(T) = 1/(\exp[\hbar\omega_\alpha/k_BT]-1)$ is the thermal occupation of the phonon mode $\alpha$ and $\langle \hat{n}_\alpha \rangle = \textsf{Tr} \{ \hat{n}_\alpha \hat{\rho}(t)\}$ denotes the quantum average of the phonon occupation operator. The later depends on the time through the {\it electron-phonon} density matrix $\hat{\rho}(t)$. Then, if $\langle \hat{n}_\alpha \rangle > \overline{n_\alpha}(T)$ during the time evolution of the system, because of the energy that is exchanged with the electron, mechanical energy flows out of the plataform to the bulk of the sample.
The relevance of other dissipation mechanisms is discussed in section \ref{clamping}.

The electron-phonon basis set is formed by the direct product of the one-electron states $|\phi_\kappa \rangle$ with the multi-phonon states $|{\bf n} \rangle \equiv |n_1,n_2,n_3,...,n_N \rangle$. Here, $n_\alpha = 0,1, \ldots, n$ denotes the number of phonon quanta per mode $\alpha$. A total of $N$ distinct phonon modes are considered, so that a typical basis vector is written as
\begin{equation}
\left|\kappa; {\bf n} \right> =
|\phi_{\kappa} \rangle \otimes \prod_{\alpha=1}^N
\frac{1}{\sqrt{n_{\alpha}!}} \, (a_{\alpha}^\dag)^{n_{\alpha}} \,
|0\rangle \ ,
\label{base}
\end{equation}
with $N = 40$ and $n_\alpha \leqslant 40$ in the following calculations. The convergence of the results were tested with respect to both the electron and phonon basis sizes. The electron-phonon basis is set up according to the prescription: first, a large ($ > 1.5\times10^5$) basis set comprised by states of  Eq.  (\ref{base}) is generated and energy sorted, in order to create a micro-canonical ensemble. The time evolution of the initial el-ph state is then carried out on a truncated basis with size ranging from $1\times10^5$ (for calculations made at $T=50$ mK) to $1.2\times10^5$ (for calculations made at $T=100$ mK). For the parameters used in the calculations, the el-ph basis is formed by the combination of 50 to 60 distinct electronic states $|\phi_\kappa \rangle$ and approximately $15\times10^3$ different $|n_1,n_2,n_3,...,n_N \rangle$ phonon states. We point out that the calculations do not converge if the electronic basis does not include several angular momentum states. We have used $7\leq|l|\leq10$, but the appropriate number depends on the characteristic energies of the system.

\subsection{Dynamics by the split-time Chebyshev method}

The time evolution of the coupled electron-phonon state $|\Psi\rangle$ is governed by the time-dependent Schr\"odinger equation 
\begin{eqnarray}
i\hbar \frac{d|\Psi\rangle}{dt} = \hat{H}|\Psi\rangle \ ,
\label{Schrodinger}
\end{eqnarray}
with $\hat{H}$ given by the hamiltonian in Eq. (\ref{hamiltonian}). 
Within a sufficiently small time slice the time dependence of the hamiltonian can be disregarded and the solution of Eq. (\ref{Schrodinger}) for a small time step $\Delta t$ is
\begin{eqnarray}
|\Psi(\Delta t)\rangle = \exp{\left(-\frac{i}{\hbar}\hat{H}\Delta t\right)}~|\Psi(0)\rangle = \hat{U}(\Delta t)~|\Psi(0)\rangle \ .
\end{eqnarray}
The time propagation of $|\Psi(t)\rangle$ has been checked for decreasing time steps in order to validate the approximation. Within a given time slice, a very efficient propagation scheme consists of expanding the evolution operator $\hat{U}$ in terms of the orthogonal Chebyshev polynomials $T_k$ \cite{Kosloff81,Dobrovitski03}
\begin{eqnarray} 
\hat{U}(t) = \lim_{K\rightarrow\infty} \sum_{k=0}^{K}c_k(t)~T_k(\hat{H}) \ ,
\end{eqnarray}
with the coefficients $c_k(t)$ calculated using the orthogonal property of the Chebyshev polynomials
\begin{eqnarray}
c_k(t) &=& \frac{2-\delta_{k}}{\pi}~\int_{-1}^{1} \frac{T_k(x)~e^{-ixt}}{\sqrt{1-x^2}}~dx \nonumber \\
&=& (2-\delta_{k})(-i)^k~J_k(t) \ ,
\end{eqnarray}
where $J_k(t)$ is the Bessel function of order $k$. Because the Chebyshev polynomials $T_k(x) = \cos{[k~\arccos(x)]}$ are defined in the interval $[-1,1]$, the hamiltonian $\hat{H}$ has to be rescaled according to
\begin{eqnarray}
\mathcal{H} = 2~\frac{(H-\overline{E})}{\epsilon} \ ,
\label{transformation}
\end{eqnarray}
where $\overline{E} = (E_{max} + E_{min})/2$ corresponds to the medium value of the eigenvalues of the original hamiltonian and $\epsilon \gtrsim E_{max} - E_{min}$ is a superior value for the range of its span, so that $-1 \leqslant \langle \kappa; {\bf n}| \mathcal{H} |\kappa; {\bf n} \rangle \leqslant 1$. 

Following transformation (\ref{transformation}), the time step is redefined as $\tau = (\epsilon \Delta t/2 \hbar)$. 
Therefore, the time evolution for the initial state $|\Psi(t_0)\rangle = |\Psi_0 \rangle$ is given by
\begin{eqnarray}
|\Psi(\Delta t)\rangle &=& 
e^{-i \mathcal{H} \tau}~e^{-i \overline{E}\Delta t/\hbar} |\Psi_0 \rangle \\
        &\approx& e^{-i \overline{E}\Delta t/\hbar}~\sum_{k=0}^{K} c_k(\tau)~|\Psi_k \rangle \label{expansion} ~,
\end{eqnarray}
with $c_k(\tau) = (2-\delta_{k}) (-i)^k J_k(\tau)$. The  vector states $|\Psi_k \rangle = T_{k}(\mathcal{H}) |\Psi_0 \rangle$ are easily obtained from the recurrence relation
\begin{eqnarray}
|\Psi_{k} \rangle = 2 \mathcal{H} |\Psi_{k-1} \rangle - |\Psi_{k-2} \rangle \ ,
\end{eqnarray}
and from the particular results $T_0(\mathcal{H}) |\Psi_0 \rangle = |\Psi_0 \rangle$ and $T_1(\mathcal{H}) |\Psi_0 \rangle = \mathcal{H} |\Psi_0 \rangle$. 

The advantage of the Chebyshev propagation method, compared to other expansion schemes is due to the fact that when $k > \tau$, $J_k(\tau)$ tends to zero exponentially fast. Therefore, expansion (\ref{expansion}) can be truncated at $K \gtrsim \tau$, with the actual value of $K$ depending on the chosen accuracy $\varepsilon$. In our calculations $\varepsilon = 5\times10^{-5}$, which yields $K/\tau \approx 1.1$ for a time step $\Delta t = 0.25~ns$. In the Chebyshev method, CPU time scales linearly with basis size and memory use is highly optimized because the matrix elements of $\mathcal{H}$ can be calculated on the fly, or stored in packed format for sparse hamiltonians. 

\section{Simulation Results}
\label{simulations}

The quantum-mechanical equations of motion of a two-level atom coupled to a single resonant mode of a cavity in the absence of dissipation were solved exactly by Jaynes and Cummings \cite{Jaynes&Cummings}. They found that this interaction leads to a simple oscillation of energy between the atom and the cavity mode.
This result corresponds to the well known semiclassical solution of a two-level system interacting with the electromagnetic field \cite{Scully}. Sachdev \cite{Sachdev} examined the quantum theory of a two-level atom in a damped cavity, in the limit of small damping, and showed that thermal fluctuations wash out the Rabi oscillations as the temperature increases. 

However, we find that a simple two-level model for the electron in the QD is not adequate to describe the present results, even if the higher energy angular momentum states are not significantly populated. The reason is that the angular momentum of the electronic system is not stationary anymore,
because of its coupling with the vibrational modes of the platform. Thus, although the initial state of the electronic system may consist of a wavepacket of well defined angular momentum, several other angular momentum states will take part in the dynamics. 

In the remainder of the paper we investigate the coherent dynamics of the electronic state of the QD coupled to a thermal field comprised by the phonon modes of the platform at the initial temperature $T$. The total vibrational energy of the multi-mode cavity state ${\bf n} \equiv (n_1,n_2,\ldots,n_N)$ is $E_{ph}({\bf n}) = \sum_\alpha \left( \hat{n}_\alpha +\frac12 \right) \hbar \omega_\alpha$, thus the density matrix elements of the phonon ensemble in a thermal field configuration are given by  
\begin{eqnarray}
 \rho^{ph}_{{\bf n n}}(0) = \frac{exp(-E_{ph}({{\bf n}})/k_B T)}{\sum_{\{{\bf n}\}} exp(-E_{ph}({\bf n})/k_B T)} \ .
\label{thermal-field}
\end{eqnarray}
In the following calculations two cases are considered: the initial temperature of the thermal field is either $T=50~mK$ or $T=100\ mK$, which correspond to typical conditions in the study of quantum coherent phenomena \cite{Elzerman,Kotthaus,Gorman,Hayashi}. 

\begin{figure}
\begin{center}
 \includegraphics[width=6.5cm]{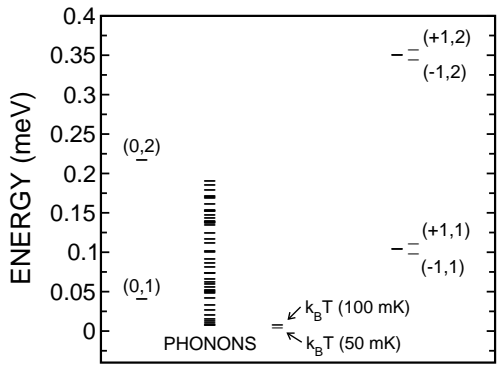}
\caption{Diagram with the energies involved in the problem: the energies $E_{l\nu}$ of the eigenstates  of $H_{el}$ (see Eq.(\ref{Zeeman})) are labeled by the angular momentum ($l$) and the radial quantum number ($\nu$); the energies of the lowest 40 phonon modes, and the thermal energy parameter ($k_BT$). For the case of a perpendicular magnetic field, the Zeeman splitting is indicated for the electronic states with $l\not=0$. The magnetic field is $B=500~G$.} 
\label{energies}
\end{center}
\end{figure}

The relevant energy scales of the system are shown in Figure \ref{energies}, where we have the phonon energies $E_{ph} \sim 8\times10^{-3} - 10^{-1}$ meV, the electronic energies $E_{el} \sim 10^{-1}$ meV, the electron-phonon matrix element energies $E_{el-ph} \sim 10^{-7} - 10^{-4}$ meV and the thermal energy $E_{th} \sim 10^{-3} - 10^{-2}$ meV. The Zeeman splitting between states with $l=\pm1$ is also shown, $\Delta E_{Zeeman} \approx 6\times10^{-3}$ meV for $B=500~G$. The quality factor of the platform is $Q=100$ for all phonon modes. The calculations are performed for a suspended platform made of crystalline silicon \cite{silicon}, with  dimensions $w=1.2~\mu m$, $l=200~nm$  and $t=50~nm$ and a circular QD of radius $R = 75~nm$. 

The el-ph dynamics is sensitive to the position of the quantum dot in the phonon cavity, due to the interplay between the distinct symmetries of the circular QD states and the rectangular phonon modes of the platform. In the following calculations, the QD is placed at the center of the platform, which is the most symmetric position.

\subsection{Intrinsic electron-phonon bath dynamics}
\label{noB}

The dissipation mechanisms can be generally understood as having external or internal origin.
Initially, we investigate the physics of the intrinsic el-ph bath relaxation dynamics in the absence of magnetic fields for the idealized case Q $\rightarrow \infty$.
Because the electronic angular momentum $L_{el}$ is stationary for the isolated QD, its time evolution is particularly useful in describing the electron-phonon interaction. Thus, we look into the time evolution of the rotational wavepacket $|\Psi(t=0)\rangle = |1,1\rangle \otimes |\chi_{bath}\rangle$, which describes the single electron QD in the first excited state ($L_{el}=1$) in contact with the phonon bath described by Eq. (\ref{thermal-field}). 

\begin{figure}[h]
\includegraphics[width=7cm]{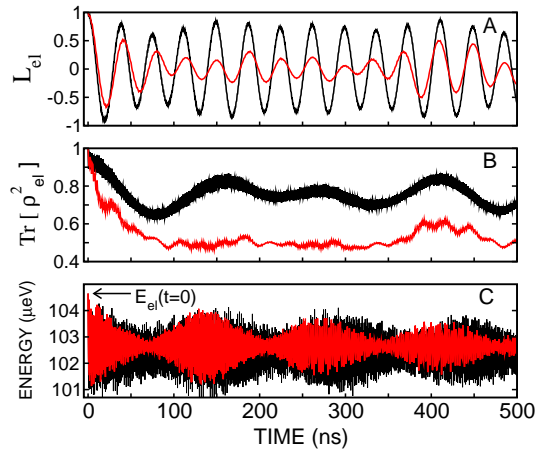}
\caption{Time evolution of the coupled electron-phonon state assuming only internal decoherence and no external mechanical dissipation. From top to bottom the graphs correspond to: A) average electronic angular momentum $L_{el}(t) = \textsf{Tr}_{el}\{\hat{L}\hat{\rho}_{el}(t)\}$, B) electronic decoherence parameter $\Gamma_{el}(t) \equiv \textsf{Tr}\{\hat{\rho}^2_{el}(t)\}$ and C) electronic energy. The curves correspond to  T = 50 mK (black) and T = 100 mK (red).}
\label{L_el}
\end{figure}

Allowing the initial state $|\Psi(0)\rangle$ to evolve according to Eq. (\ref{expansion}), Figure \ref{L_el}-A shows the electronic angular momentum $L_{el}(t) = \textsf{Tr}\{\hat{\rho}_{el}(t)\hat{L}\}$ as it oscillates in time between the degenerate states $L_{el} = \pm 1$, due to the interaction with the phonon bath in the platform. $\hat{\rho}_{el}(t) = \textsf{Tr}_{ph}\{\hat{\rho}(t)\}$ is the reduced electronic density matrix and $\textsf{Tr}_{ph}$ designates the trace over the phonon states. The black curves represent the calculations made for the initial phonon temperature  $T=50$ mK and the red curves correspond to $T=100$ mK. 

Rabi's theory is useful in interpreting the general features of the  numerical results and can be used to determine the effective electron-phonon coupling parameter $\overline{\gamma}_{ep}$ responsible for the $L_{el}$ oscillations. Analysing the dynamics of the simpler case, {\it i.e.} $T=50$ mK in Fig. \ref{L_el}-A, we observe that $L_{el}$ oscillates with almost constant amplitude between the values -1 and +1. In this case, $L_{el}$ is equivalent to the inversion amplitude $W(t) = \sum_{{\bf n}} \left(|C_{+1,{\bf n}}|^2 - |C_{-1,{\bf n}}|^2 \right)$, where $|\Psi(t)\rangle = \sum_{\kappa,{\bf n}} C_{\kappa,{\bf n}}|\kappa,{\bf n} \rangle$. Moreover, for $T=50$ mK, $\langle n_\alpha \rangle \ll 1$ for all phonon modes and one can assume the vacuum state for the phonon cavity. Therefore, $L_{el}$ can be approximately described by
\begin{eqnarray}
L_{el}(t) = \frac{\omega_{\pm 1}^2 + \left(\frac{2 \overline{\gamma}_{ep}}{\hbar}\right)^2 \cos{\left[\sqrt{\omega_{\pm 1}^2+\left(\frac{2\overline{\gamma}_{ep}}{\hbar}\right)^2}t\right]}}
{\omega_{\pm 1}^2 + (2\overline{\gamma}_{ep}/\hbar)^2}  \ ,
\label{Rabi1}
\end{eqnarray}
where $\omega_{\pm 1} = (E_{1,1} - E_{-1,1})/\hbar$.
In the absence of magnetic fields $\omega_{\pm 1} = 0$ and Eq. (\ref{Rabi1}) simplifies to $L_{el} = cos(2\overline{\gamma}_{ep} t/\hbar)$.  The evaluation of the effective electron-phonon coupling yields $\overline{\gamma}_{ep} \approx 5.10^{-5}~meV$ for this simple case, which is consistent with the calculated matrix elements $g_{\kappa' \alpha \kappa}$ in Eq. \ref{hamiltonian}. 
It is important to notice that the dynamics of the el-ph system is caused mainly by virtual phonon transitions.
When a weak magnetic field is applied, the electron-phonon matrix elements do not change, but the degeneracy condition is broken and $\omega_{\pm 1} = 2E_{Zeeman}/\hbar$. Then, one expects that the oscillation amplitude of $L_{el}$ decreases. Real transitions can also occur as a result of accidental resonances involving the $l=0$ state and high energy phonon modes. Those effects are discussed in section \ref{withB} along with the simulation results. 

A variety of different behaviors is observed for $L_{el}$, just by changing the edge conditions of the platform or by displacing the QD from its center. The effects include the plain oscillatory dynamics, beatings and the overdamped decay of $L_{el}$. The asymptotic value of the electronic angular momentum for all the cases is $L_{el}=0$. That corresponds to the equal occupation of the electronic states with $l=\pm1$, a small occupation of states with $l=0$ and a very small occupation of the other $l$ states.

As expected, the damping of the coupled electron-phonon oscillations varies with the initial bath temperature. 
To evaluate more precisely the degree of decoherence in the electronic wavefunction ({\it i.e.}, the rotational wavepacket) we calculate the purity of the electronic state $\Gamma_{el}(t) \equiv \textsf{Tr}\{\hat{\rho}^2_{el}(t)\}$.
At time $t=0$ the electronic system is in the pure state described by $|\Psi(t=0)\rangle$ and $\Gamma_{el} = 1$. The electronic decoherence is presented in Figure \ref{L_el}-B, for the initial phonon temperatures $T=50$ mK (black) and $T=100$ mK (red). The interaction with the phonon modes of the platform is responsible for the decay of $\Gamma_{el}$, which saturates at $\Gamma_{el} \approx 1/2$, evidencing the fact that the phases of the wavepacket evolve towards an even distribution among the $\left|\kappa; {\bf n} \right>$ states with angular momenta $l=\pm 1$. The calculations show that the internal el-ph friction causes electronic decoherence in less than 100 ns for $T=100$ mK.
We point out, however, that such time scale for the electronic decoherence should be understood as a higher bound , since the external decohering mechanisms, like the  clamping  loss, have not yet been taken into account.

By examining the autocorrelation of the entire electron-phonon bath state, $\xi(t) = \langle \Psi(0)|\Psi(t) \rangle$, we have evidenced for all the cases investigated that the 
electron-bath system undergoes a fast phase decoherence, within the time scale $\tau_{el-ph}\approx 0.1$ ns. It is followed by the electronic decoherence, evinced by the decay of $\Gamma_{el}$ within $\tau_{el}\approx 100$ ns,  and by an even  slower internal energy dissipation process (el-ph friction)  that leads to the formation of the polaron state in the QD. 
 The electronic energy dissipation is shown in Fig \ref{L_el}-C, a stationary energy distribution is reached within the microsecond time scale ($\tau_{erg}$).
The initial energy of the electron, indicated by the arrow, is much higher than the thermal energy of the phonon bath, leading to a recurrent energy exchange between the electron and the phonon modes of the platform. Energy is exchanged between the electron and the phonon bath primarily through the $l=0$ state and the high energy phonon modes ($\alpha$ = 15, 16, 17). 
For the sake of comparison, similar beating effects have been observed in molecular dynamics simulations of nanomechanical energy exchange between single-walled carbon nanotubes \cite{Greaney}.

\subsection{Dissipation effects}
\label{clamping}

Energy dissipation is a central issue for the operation of nano- and microelectromechanical structures (MEMS).
The identification of the source of mechanical energy dissipation is, nonetheless, a complex problem because different mechanisms can be determining in distinct physical situations \cite{Qfactor,Simon}.
Usually, the investigation is focused on the relaxation of energy from a driven resonant mode of a cantilever, bridge or structure alike. 
For a suspended NEMS operating in vacuum, vibrational energy is radiated out of the structure through the resonator attachments \cite{Jimbo,Cross,Photiadis,HuangNJP,Imboden}. On the other hand, intrinsic dissipation also contributes to take energy out of the resonant mode. Such mechanisms include phonon-phonon and electron-phonon scattering, thermoelastic effect \cite{Lifshitz}, surface and bulk defects \cite{Qfactor,Mohanty02,Zolfagharkhani}.

At sub-Kelvin temperatures and in the mesoscopic regime, phonon-phonon scattering and thermoelastic relaxation can be safely disregarded \cite{Lifshitz} in comparison to the other intrinsic friction mechanisms. Dissipation due to defects, however, can not in principle be discarded. Defects arise most frequently from broken and dangling bonds on the surface of the structure and from contamination by other atoms. Zolfagharkhani and collaborators \cite{Zolfagharkhani} have conducted a detailed investigation of the quantum friction in nanomechanical oscillators at millikelvin temperatures. They showed, in accordance with other studies \cite{Mohanty02}, that dissipation in this regime is dominated by the interaction of the resonant mechanical modes with localized defects. They also found that dissipation decreases with temperature, but saturates at $Q^{-1} \approx 3\cdot10^{-5}$ for $T\lesssim 100$ mK, in the single-crystal Si nanobridges investigated.

Undoubtedly, the dominant dissipation mechanism in the platform is attachment loss, because of its wide clamping edge. Throughout the literature, experimental and theoretical investigations concentrate on cantilever and bridge geometries, owing to the fact that the attachment losses can be minimized for high aspect ratio structures ({\it} i.e., $l \gg w$) \cite{HuangNJP,Imboden}. 
Therefore, it is difficult to estimate the quality factor of the suspended platform.
However, two different theoretical expressions yield the same estimate for the $Q$ of our model structure. According to the expression of Photiadis and Judge \cite{Photiadis} $Q_{PJ} \approx 3.2~(l^5/wt^4) \approx 137$ whereas the estimate due to Jimbo and Itao \cite{Jimbo} yields $Q_{JI} \approx 2.17~(l/t)^3 \approx 139$. Both estimates are in qualitative agreement with the experimental results of Huang and collaborators \cite{HuangNJP}, obtained for a series of SiC nanobridges. That work measured a quality factor $Q$ = 500, for a wide ($w = 120$ nm) and short ($1 = 1~\mu m$) doubly clamped beam, with fundamental resonant frequency $f_1 \approx 1$ GHz. 

Therefore, we estimate that the overall quality factor of the platform of Fig. \ref{structure} is $Q \approx$ 100.  Assuming that the structure has $Q$=100, one finds that the lifetime of a resonant mechanical mode with frequency $f$ = 1.5 GHz
is limited by the losses in the platform to $\tau_{_Q} \approx Q/2\pi f \approx 10$ ns. However, the dynamics of the el-ph system is caused by virtual phonon transitions that do not result in direct energy transfer to the phonon system, nor to a particular resonant phonon mode. In that case, one expects that the electronic oscillations should persist even in the presence of mechanical dissipation.
\begin{figure}
 \includegraphics[width=7cm]{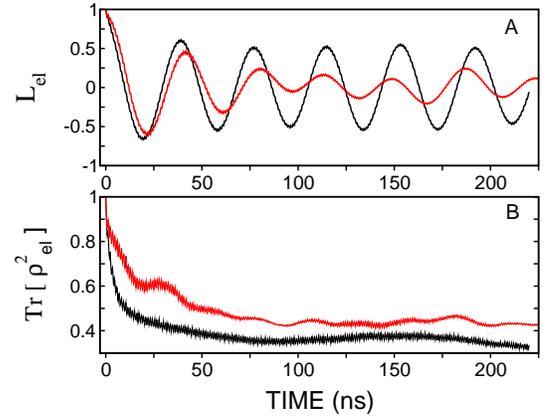}
\caption{Time evolution of the coupled electron-phonon dynamics, including attachment loss dissipation with Q = 100. A) average electronic angular momentum $L_{el}(t) = \textsf{Tr}_{el}\{\hat{L}\hat{\rho}_{el}(t)\}$, B) electronic purity $\Gamma_{el}(t) \equiv \textsf{Tr}\{\hat{\rho}^2_{el}(t)\}$. The curves correspond to  T = 50 mK (black) and T = 100 mK (red).}
\label{dissipation}
\end{figure}

Since the phonon system is described by a thermal field from the outset,
the dissipation of the mechanical energy of the platform is described within the relaxation time approximation in hamiltonian (\ref{hamiltonian}), with $Q$ = 100 for all phonon modes. According to $\hat{H}_{relax}$, mechanical energy of a given phonon mode is radiated out to the bulk of the sample  if $\langle n_\alpha \rangle > \overline{n_\alpha}(T)$, and vice-versa. As evinced by Fig. \ref{dissipation}, the mechanical dissipation produces a strong effect on the low temperature el-ph bath dynamics ($T=50$ mK) whereas the $T=100$ mK case is just weakly affected. 
The reason is that a small amount of energy that is transferred to the low temperature phonon system renders enough to drive that system out of the equilibrium.
The result also indicates that electron-phonon friction becomes the main source of electronic decoherence for $T > 100$ mK, but below that temperature clamping loss is determining. 

\subsection{Dynamics of $L_{el}$ in the Bloch sphere}

The entanglement between the orbital angular momentum states $|l=\pm1,\nu\rangle$ 
is revealed through the analysis of the el-ph hamiltonian matrix $\hat{H}_{el-ph}$. It has been
noted that, for a circular QD located at the center of the platform, the matrix representation of the total hamiltonian $\hat{H}$ can be written in blocks of fixed $l$'s as $\mathbb{H}_{l'l} = \mathbb{A}_{l'l} +
(i)^{\mbox{\scriptsize{sign}}(l' - l)} \, \mathbb{B}_{l'l}$, where $\mathbb{A}$ and $\mathbb{B}$ are real, symmetric and mutually disjoint matrices.
Moreover, the matrix $\mathbb{A}$ ($\mathbb{B}$) is responsible for the coupling between the electron and the even (odd) parity phonon modes.
Therefore, starting from $|\Psi(t=0)\rangle = |1,1\rangle \otimes |\chi_{bath}\rangle$, the $\mathbb{A}$ matrix operates like the generator of rotations $\hat{R}_x$, leading the electronic state through a precessional movement around the $\hat{e}_x$ axis of the angular momentum space. Likewise, the $\mathbb{B}$ matrix functions like the $\hat{R}_y$ operator, rotating the electron state around the $\hat{e}_y$ axis.

\begin{figure}
 \includegraphics[width=7cm]{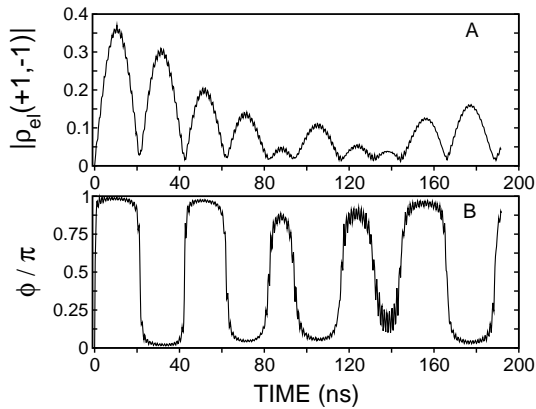}
\caption{Time evolution of the density matrix element $\rho_{el}(1,-1)$ for T = 100 mK and Q = 100. A) the modulus $|\rho_{el}(1,-1)|$ and B) the phase normalized by $\pi$.}
\label{rho_el}
\end{figure}
 
Since the time evolution of $|\Psi(0)\rangle$ results mainly in occupation of the $l=\pm 1$ states with $\nu=1$ we can, for the sake of clarity, restrict ourselves to that subspace and describe the electronic state in the Bloch sphere. 
An the arbitrary state in the Bloch sphere is written as $|\psi\rangle = \cos{\frac{\theta}{2}}|+1\rangle_z + \exp{(i\phi)}\sin{\frac{\theta}{2}}|-1\rangle_z$, with $0<\theta<\pi$ and $0<\phi<2\pi$.
In the current model the dynamics of $|\Psi(t)\rangle$ in the Bloch sphere
is evidenced by the density matrix element $\rho_{el}(+1,-1)=\rho^\ast_{el}(-1,+1)$, with $\nu=1$, whose modulus ($|\rho_{el}(+1,-1)|$) and phase ($\phi$) are shown in Figure \ref{rho_el} for $T=100$ mK and Q=100. 
The numerical calculations show that the electron state undergoes a precessional movement around the $\hat{e}_y$ axis, that is, as $L_{el}(t)$ oscillates between $|L_z\rangle = |\pm 1\rangle_z$ (Figs. \ref{L_el} and \ref{dissipation}) it passes through $|L_x\rangle = |\pm 1\rangle_x = \left(|+1\rangle_z \pm |-1\rangle_z \right)/\sqrt{2}$. This coherent dynamics persists during the relaxation time $\tau_{el} \approx 100$ ns and after that the rotational wavepacket looses most of the coherence. Furthermore, it evinces that the el-ph coupling with the odd parity phonon modes is the main responsible for the observed Rabi oscillations. 
Finnaly, we point out that 
the orbital angular momentum of the suspended QD system behaves like a spin 1/2 system and the interaction with the phonon modes of the platform play the role of a transverse magnetic field.

\subsection{Electron-phonon dynamics in the presence of a weak magnetic field}
\label{withB}

\begin{figure}
 \includegraphics[width=7cm]{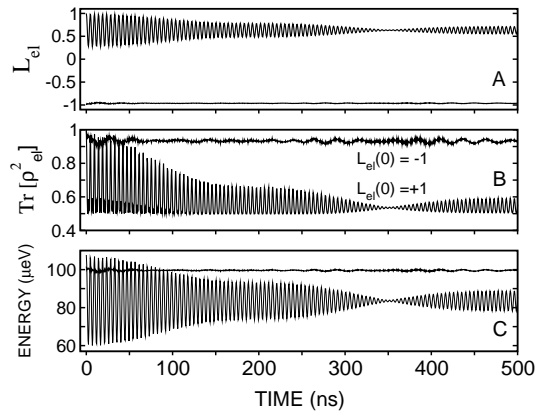}
\caption{Time evolution of the coupled electron-phonon state in the presence of a weak magnetic field, assuming only internal decoherence and no external mechanical dissipation. From top to bottom the graphs correspond to: A) average electronic angular momentum $L_{el}(t) = \textsf{Tr}_{el}\{\hat{L}\hat{\rho}_{el}(t)\}$, B) electronic purity $\Gamma_{el}(t) \equiv \textsf{Tr}\{\hat{\rho}^2_{el}(t)\}$ and C) electronic energy. The curves correspond to initial states with $L_{el}(t=0) = \pm 1$ and T = 100 mK. The magnetic field is $B=500~G$.}
\label{L_el_B}
\end{figure}

The dynamics of the coupled electron-phonon system is substantially changed in the presence of a magnetic field $\vec{B}$ applied perpendicularly to the plane of the QD. As noted in section \ref{model-A}, the diamagnetic energy term in Eq. (\ref{magnetic2}) can be disregarded if the magnetic field is sufficiently weak. To satisfy the weak field condition, we set $B=500~G$, so that $E_{diam}/E_{Zeeman} = 0.1$ and $E_{diam}/E_{_{GS}} = 0.008$, where $E_{_{GS}}$ is the energy of the electronic ground state $|\kappa\rangle = |0,1\rangle$. In this regime the electron states with $\pm l$ are Zeeman splitted but the electronic wavefunctions remain unaltered. Thus, the ensuing effects have a purely kinetic origin, since they arise exclusively because of the elimination of the $\pm l$ degeneracy.
		
Figure \ref{L_el_B} presents the dynamics of the electron-phonon system in the presence of $B$, disregarding mechanical losses for the moment, at the initial bath temperature $T=100$ mK. For the sake of completeness, we also show the evolution of the initial state $|\Psi(0)\rangle$ with $L_{el}(0)=-1$. The physics evinced in Fig. \ref{L_el_B} should be contrasted with that presented in Fig. \ref{L_el}. 

Two different physical effects occur for $L_{el}(0)=\pm 1$, because of the Zeeman splitting of the initial state. In the case $L_{el}(0)=-1$ the oscillations of $L_{el}$ are almost completely eliminated and $\Gamma_{el}(t) \approx 1$ for all times. The reason is that $0 < \omega_{\pm 1} < \omega_{\alpha =1}$, rendering the state $l=-1$ decoupled from the rest of the system. As the magnetic field decreases, the oscillations of $L_{el}$ quickly build up to the original value. The magnetic field also reduces the effect of the temperature in the el-ph relaxation, so that the electronic decoherence and the energy dissipation processes are delayed, Figs. \ref{L_el_B}-B and C.

\begin{figure}
 \includegraphics[width=7cm]{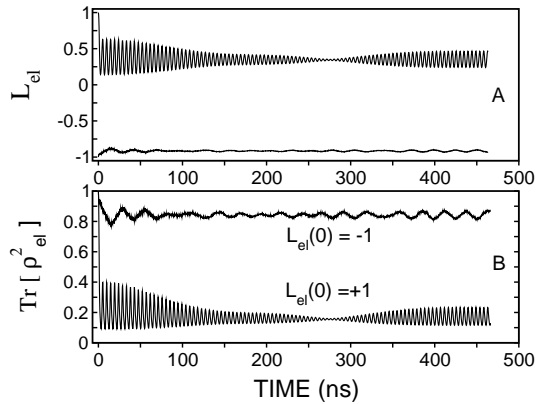}
\caption{Time evolution of the coupled electron-phonon state in the presence of a weak magnetic field, including attachment loss dissipation with Q = 100. A) average electronic angular momentum $L_{el}(t) = \textsf{Tr}_{el}\{\hat{L}\hat{\rho}_{el}(t)\}$ and B) electronic purity $\Gamma_{el}(t) \equiv \textsf{Tr}\{\hat{\rho}^2_{el}(t)\}$. The curves correspond to initial states with $L_{el}(t=0) = \pm 1$ and T = 100 mK. The magnetic field is $B=500~G$.}
\label{dissipation-B}
\end{figure}

For the case $L_{el}(0) = +1$, in addition to the decoupling $0 < \omega_{\pm 1} < \omega_{\alpha =1}$, the Zeeman splitting produces the accidental resonance $E(1,1)-E(0,1) \simeq \hbar\omega_{\alpha=17}$. That resonance sets up a strong Rabi oscillation between the $l=+1$ and the $l=0$ electronic states, which is responsible for the oscillation of $L_{el}$ and, specially, the intense energy exchange with the phonon mode $\alpha=17$. The resonant effect is very sensitive to the detuning $\delta$. For the calculations presented in Fig. \ref{L_el_B} we find that $\delta=0.004\times\omega_{\alpha=17}$. However, by setting $\delta=0.02\times\omega_{\alpha=17}$ the oscillation amplitude of $L_{el}$ decreases to a third.

 For a QD located at the center of the platform, the phonon modes of odd parity are  the ones that generally couple more strongly with the electronic states.

The mechanical dissipation affects the el-ph dynamics of the $L_{el} = \pm 1$ cases differently, as shown in Figure \ref{dissipation-B}. Because the state with $L_{el}(0) = -1$ exchanges very little energy with the phonon bath, the latter remains close to the thermal equilibrium and the attachment loss produces a minor dissipation effect. 
The $L_{el}(0) = +1$ state, otherwise, exchanges a large amount of energy with the high energy phonon modes of the bath (Fig. \ref{L_el_B}-C). Consequently, for those modes $\langle n_\alpha \rangle \gg \overline{n_\alpha}(T)$, which drives the phonon system out of equilibrium and leads to a strong mechanical energy dissipation.

\section{Conclusions}
\label{conclusions}

We have investigated the quantum dynamics of the electron-phonon system for an unconventional design of nanoelectromechanical resonator. The interplay between well defined orbital angular momentum states of a circular QD and the phonon modes of a wide nanomechanical resonator (here called platform) gives rise to a coherent electron-phonon dynamics. The phonon ensemble is comprised by the quantized vibrational modes of the suspended platform and is described by a thermal field. 
Virtual phonon transitions cause the electronic state to undergo Rabi oscillations between the degenerate angular momentum states $l=\pm1$, without direct energy transfer to the phonon ensemble. That fact circumvents the dissipation effects caused by the mechanical (attachment) losses in the nanoresonator. By analysing the electronic density matrix, it is evidenced that the orbital angular momentum of the QD behaves like a spin 1/2 system in the presence of a transverse magnetic field. It is also shown that a weak magnetic field can be used to suspend the dynamics of the electron-phonon system, by decoupling the angular momentum states $\pm l$ because of the Zeeman splitting. Several energy loss mechanisms are considered for the platform and the attachment loss is regarded as the main source of dissipation.

The results cast light upon the underlying physics of a system that behaves like a single-QD charge-qubit, with the orbital angular momentum of the electron as the quantum bit variable. 
It differs from the usual charge-qubit structures, which consist of two adjacent quantum dots coupled by a tunneling mechanism, as well as from the spin-qubit devices. The proposed structure, nevertheless, has characteristics in common to both original concepts, since the orbital angular momentum of the QD plays the role of the spin.  

Despite the interesting properties, the proposed NEMS poses significant challenges compared to the usual cantilever and bridge structures. It is, therefore, desirable to consider alternative geometries for the platform, with the purpose of minimizing the attachment losses without removing the electronic oscillations. The current model evidences the prototypical characteristics of the structure. 

\section*{Acknowledgments}

The author is grateful to a research fellowship from CNPq/Brazil, as well as to a generous allocation of computer time from NACAD/COPPE and CENAPAD/Campinas in Brazil.


\begin{thebibliography}{99}
\bibitem{silicon} In the calculations we use $C_{DP} = 9.8\times10^3~meV$ and $m_e=0.95m_0$ for the in-plane effective mass of the electron.
\end{thebibliography}
\end{document}